# Revisiting the dynamics of *finite-sized* satellite near the planet in ER3BP

**Sergey Ershkov\***, Affiliation[1]: Plekhanov Russian University of Economics,

Scopus number 60030998, e-mail: sergej-ershkov@yandex.ru

Affiliation[2]: Sternberg Astronomical Institute, M.V. Lomonosov's Moscow State

University, 13 Universitetskij prospect, Moscow 119992, Russia,

**Dmytro Leshchenko**, Odessa State Academy of Civil Engineering

and Architecture, Odessa, Ukraine, e-mail: leshchenko_d@ukr.net,

**Alla Rachinskaya**, Odessa I. I. Mechnikov National University,

2 Dvoryanskaya St., Odessa, Ukraine, e-mail: rachinskaya@onu.edu.ua

## Abstract

A novel approach for solving equations of motion of *finite-sized* satellite supposed to be moving in a proximity and around the planet in the elliptic restricted three-body problem, ER3BP is presented in this semi-analytical investigation. We consider two primaries, $M_{Sun}$ and $m_{planet}$ (the last is secondary in that binary system), both are orbiting around their barycenter on *elliptic* orbits. Satellite is considered to be the solid ellipsoid having nearly spherical form, with its gravitational potential to be given by a formula of MacCullagh type.

Our aim is to revisit previously presented in work [Ashenberg, 1996] approach and to investigate the updated type of the satellite dynamics correlated implicitly to a kind of *trapped* motion (in the synodic co-rotating Cartesian coordinate system) in so way that satellite will always to be located near the secondary planet, $m_{planet}$, moving on quasi-stable elliptic orbit.

# 1. Introduction.

The planar dynamics of satellite of *finite size* has been presented previously in case of elliptical restricted three-body problem ER3BP in a seminal work of J.Ashenberg [1] (where satellite was assumed to be moving under the influence of gravitational forces of two primaries $M_{Sun}$ and $m_{planet}$, $m_{planet} < M_{Sun}$, which are orbiting around their barycenter in mutual Keplerian motion) with help of clear mathematical formulae by the system of two coupled ordinary differential equations of 2nd order. *Videlicet*, these equations are supposed to be describing the *translational* motion of satellite (i.e., equation of momentum) in pulsating coordinate system, whereas the second equation was introduced by him to describe the *planar rotational* motion of satellite [1]. We will investigate in the current research only the *translational* motion of satellite. It is worth to note that a lot of outstanding fundamental works were devoted to the analytical or numerical findings with respect to the approximated approaches for solving the same problems (see for example the profound work of Beletsky [2] with respect to the *planar rotational* motion of satellite, but not limited to).

In the ER3BP, the positions of the primaries are not the same for the reason of their permanent motion along the elliptical orbits, so that their relative distance $\rho$ is not invariable in time during the motion [3-5]

$$\rho = \frac{a_p(1-e^2)}{1+e\cos f}$$

where, $a_p$ is the semimajor axis of *elliptic* orbits of the rotating primaries around their barycenter (here, scale of distances is chosen so that $\{a_p(1-e^2)\} = 1$), $e$ is the eccentricity, $f$ is the true anomaly. Besides, angular motion is given by

$$\frac{df}{dt} = \left(\frac{GM}{a_p^3(1-e^2)^3}\right)^{\frac{1}{2}} (1+e\cos f)^2 \qquad (1)$$

where, $G$ is the Gaussian constant of gravitation law, $M$ is the sum of masses of primaries, the unit of time is chosen so that constant *G* equals to 1.

According to [1], equations of the planar dynamics of satellite of *finite size* have been derived analytically in case of satellite's motion in the vicinity of equilibrium points {$L_1$, ..., $L_5$} for elliptical restricted three-body problem ER3BP (regarding *collinear* points {$L_1$, $L_2$, $L_3$}, see works [6-8]; as for the triangular points {$L_4$, $L_5$}, see works [9-11]). We will concentrate our efforts on exploring the *planar* dynamics of satellite of *finite size* for the case of solid ellipsoid having nearly spherical form, with its gravitational potential to be given by a formula of MacCullagh type [3, see p. 111] in the current research. Namely, if a solid uniform ellipsoid of mass *m* is nearly spherical and has axes *a*, $\sqrt{a^2-h}$ and $\sqrt{a^2-k}$, the potential at external point $\vec{r} = \{x, y, z\}$ is

$$V(r) = -\frac{Gm}{r} - \frac{Gm}{10r^5}\left\{x^2(h+k) + y^2(-2h+k) + z^2(h-2k)\right\} \quad (2)$$

to the first order of small quantities, where $z = 0$ for the case of planar motion.

According to [1], let us present the system of equations in the scaled, pulsating, planar coordinate system $\vec{r} = \{x, y\}$ (in elliptical restricted three-body problem, ER3BP, *at given initial conditions*)

$$\ddot{x} - 2\dot{f}\dot{y} = (\dot{f})^2 x + \ddot{f} y + \frac{1}{m}\left(\frac{\partial U}{\partial x}\right),$$

$$\ddot{y} + 2\dot{f}\dot{x} = (\dot{f})^2 y - \ddot{f} x + \frac{1}{m}\left(\frac{\partial U}{\partial y}\right), \quad (3)$$

where true anomaly *f* is *the dependent* variable, $f = f(t)$ (which is the angular distance of the radius vector from pericenter of the orbit) determined by Eq. (1), whereas $r_i$ (*i* = 1, 2) are the distances of small mass *m* from each of primaries with mass *M* $_{Sun}$ and *m* $_{planet}$, accordingly [12-13]:

$$r_1^2 = (x-\mu)^2 + y^2,$$

$$r_2^2 = (x-\mu+1)^2 + y^2, \qquad (4)$$

$$U = -V(r_1) - V(r_2),$$

besides, in (3) dot indicates derivative with respect to *t*, *U* is the scalar function.

Now, the unit of mass is chosen in (3) so that the sum of the primary masses *M* in (1) is equal to 1. We suppose that *M Sun* ≅ *1 - μ* and *m planet* = *μ*, where *μ* is the ratio of the mass of the smaller primary to the total mass of the primaries and $0 < \mu \leq 1/2$. We neglect by non-classical effects considered in works [14-16] for ER3BP.

## 2. Ways of analytical or semi-analytical solving the system of Eqs. (1)-(4).

Aiming the constructing semi-analytical solution, let us present Eqns. (2)-(4) in a form congruent for our analysis below in regard to coordinates {*x*, *y*}

$$\ddot{x} - 2\dot{f}\dot{y} = (\dot{f})^2 x + \ddot{f} y - \frac{(x-\mu)}{\left((x-\mu)^2 + y^2\right)^{\frac{3}{2}}} - \frac{(x-\mu+1)}{\left((x-\mu+1)^2 + y^2\right)^{\frac{3}{2}}} +$$

$$+ \frac{\left[2x(h+k)\left((x-\mu)^2 + y^2\right)^{\frac{5}{2}} - 5(x-\mu)\left((x-\mu)^2 + y^2\right)^{\frac{3}{2}}\left\{x^2(h+k) + y^2(-2h+k)\right\}\right]}{10\left((x-\mu)^2 + y^2\right)^5} +$$

$$+ \frac{\left[2x(h+k)\left((x-\mu+1)^2 + y^2\right)^{\frac{5}{2}} - 5(x-\mu+1)\left((x-\mu+1)^2 + y^2\right)^{\frac{3}{2}}\left\{x^2(h+k) + y^2(-2h+k)\right\}\right]}{10\left((x-\mu+1)^2 + y^2\right)^5},$$

(5)

$$\ddot{y} + 2\dot{f}\dot{x} = (\dot{f})^2 y - \ddot{f} x - \frac{y}{\left((x-\mu)^2 + y^2\right)^{\frac{3}{2}}} - \frac{y}{\left((x-\mu+1)^2 + y^2\right)^{\frac{3}{2}}} +$$

$$+ \frac{\left[2y(-2h+k)\left((x-\mu)^2 + y^2\right)^{\frac{5}{2}} - 5y\left((x-\mu)^2 + y^2\right)^{\frac{3}{2}}\left\{x^2(h+k) + y^2(-2h+k)\right\}\right]}{10\left((x-\mu)^2 + y^2\right)^5} +$$

$$+ \frac{\left[2y(-2h+k)\left((x-\mu+1)^2 + y^2\right)^{\frac{5}{2}} - 5y\left((x-\mu+1)^2 + y^2\right)^{\frac{3}{2}}\left\{x^2(h+k) + y^2(-2h+k)\right\}\right]}{10\left((x-\mu+1)^2 + y^2\right)^5},$$

where we can express in (6) with help of (1)

$$\frac{df}{dt} = (1+e\cos f)^2 \Rightarrow \frac{d^2 f}{dt^2} = \frac{d\left(\frac{df}{dt}\right)}{df}\frac{df}{dt} = -2e(1+e\cos f)^3 \sin f$$

$$\dot{x} = \frac{dx}{df}(1+e\cos f)^2, \quad \dot{y} = \frac{dy}{df}(1+e\cos f)^2, \qquad (6)$$

$$\ddot{x} = \frac{d\left(\frac{dx}{df}(1+e\cos f)^2\right)}{df}(1+e\cos f)^2 = \frac{d^2 x}{df^2}(1+e\cos f)^4 - 2e(1+e\cos f)^3 \sin f \frac{dx}{df},$$

$$\ddot{y} = \frac{d^2 y}{df^2}(1+e\cos f)^4 - 2e(1+e\cos f)^3 \sin f \frac{dy}{df}.$$

Equations (5) could be simplified if we consider partial case of solutions $k = 2h$

$$\frac{d^2 x}{df^2}(1+e\cos f)^4 - 2e(1+e\cos f)^3 (\sin f)\frac{dx}{df} - 2(1+e\cos f)^4 \frac{dy}{df} =$$

$$= (1+e\cos f)^4 x - 2e(1+e\cos f)^3 (\sin f) y - \frac{(x-\mu)}{\left((x-\mu)^2 + y^2\right)^{\frac{3}{2}}} - \frac{(x-\mu+1)}{\left((x-\mu+1)^2 + y^2\right)^{\frac{3}{2}}} +$$

$$+ \frac{3h\,x\left[2\left((x-\mu)^2 + y^2\right) - 5(x-\mu)x\right]}{10\left((x-\mu)^2 + y^2\right)^{\frac{7}{2}}} + \frac{3h\,x\left[2\left((x-\mu+1)^2 + y^2\right) - 5(x-\mu+1)x\right]}{10\left((x-\mu+1)^2 + y^2\right)^{\frac{7}{2}}},$$

(7)

$$\frac{d^2 y}{df^2}(1+e\cos f)^4 - 2e(1+e\cos f)^3 (\sin f)\frac{dy}{df} + 2(1+e\cos f)^4 \frac{dx}{df} =$$

$$= (1+e\cos f)^4 y + 2e(1+e\cos f)^3 (\sin f) x - \frac{y}{\left((x-\mu)^2 + y^2\right)^{\frac{3}{2}}} - \frac{y}{\left((x-\mu+1)^2 + y^2\right)^{\frac{3}{2}}} +$$

$$- \frac{3h\,y\,x^2}{2\left((x-\mu)^2 + y^2\right)^{\frac{7}{2}}} - \frac{3h\,y\,x^2}{2\left((x-\mu+1)^2 + y^2\right)^{\frac{7}{2}}}.$$

## 3. Congruent form of semi-analytical presentation of Eqns. (7) for further solving procedure.

One of obvious ways of semi-analytical solving the system of Eqns. (5) accomplished together with Eq. (1) is to assume that solutions $\vec{r} = \{x, y\}$ of (7) belong to the class of *trapped motions* of small mass $m$ (close to the planet $m_{planet}$):

$$\frac{|\vec{r}_2|}{|\vec{r}_1|} = \frac{\left((x-\mu+1)^2 + y^2\right)^{\frac{1}{2}}}{\left((x-\mu)^2 + y^2\right)^{\frac{1}{2}}} << 1, \quad |\vec{r}_1| \cong \frac{a_p}{1+e\cos f} + \delta, \quad |\delta| << a_p \quad (8)$$

where, $|\vec{r}_2| > 10 R_p$ ($R_p$ is the radius of planet $m_{planet}$). It means that the distance of small mass $m$ from second primary should exceed the level of minimal distances out of *double* Roche-limit for this chosen primary [5]. Such an assumption (8) above should simplify the equations (7) accordingly:

$$\frac{d^2 x}{d f^2}(1+e\cos f)^4 - 2e(1+e\cos f)^3 (\sin f)\frac{d x}{d f} - 2(1+e\cos f)^4 \frac{d y}{d f} =$$

$$= (1+e\cos f)^4 x - 2e(1+e\cos f)^3 (\sin f) y - \frac{(x-\mu+1)}{\left((x-\mu+1)^2 + y^2\right)^{\frac{3}{2}}} +$$

$$+ \frac{3h x\left[2\left((x-\mu+1)^2 + y^2\right) - 5(x-\mu+1)x\right]}{10\left((x-\mu+1)^2 + y^2\right)^{\frac{7}{2}}},$$

(9)

$$\frac{d^2 y}{d f^2}(1+e\cos f)^4 - 2e(1+e\cos f)^3 (\sin f)\frac{d y}{d f} + 2(1+e\cos f)^4 \frac{d x}{d f} =$$

$$= (1+e\cos f)^4 y + 2e(1+e\cos f)^3 (\sin f) x - \frac{y}{\left((x-\mu+1)^2 + y^2\right)^{\frac{3}{2}}} - \frac{3h y x^2}{2\left((x-\mu+1)^2 + y^2\right)^{\frac{7}{2}}}.$$

## 4. Graphical plots for approximate solution and numerical findings for semi-analytical result (5).

Furthermore, we should present next the schematically imagined appropriate graphical plots for approximate semi-analytical solutions of a type (9) at Figs.1-8 below. First, we should note that dynamical character of numerical solutions and their stability (with respect to the true anomaly $f$) strongly depend on the parameter $h$ in (9) which stems from deviation of the form of satellite from ideal spheroid.

The second, we can establish the restrictions (10) which should be valid for all the possible range of meanings of such the parameter $h$

$$\frac{3h\,x^2}{10(x^2+y^2)} \ll \frac{|\vec{r}_2|}{|\vec{r}_1|} = \frac{\left((x-\mu+1)^2 + y^2\right)^{\frac{1}{2}}}{\left((x-\mu)^2 + y^2\right)^{\frac{1}{2}}} \ll 1 \qquad (10)$$

Bearing (10) in mind, let us consider the case of two primaries for system "Jupiter-Sun" where eccentricity is chosen to be $e = 0.048775$, $\mu \cong 954.509 \times 10^{-6}$ in the scheme of our numercial experiments presented on Figs. 1-4 (as for estimation of $h$ in this case, see **Appendix, A1**). We should note that we have used for calculating the data the Runge–Kutta fourth-order method with step 0.001 starting from initial values (Figs. 1-4) for the case of Jupiter:

$$x_0 = -1.007,\ (\dot{x})_0 = 0.06,\ y_0 = -0.074,\ (\dot{y})_0 = -0.33.$$

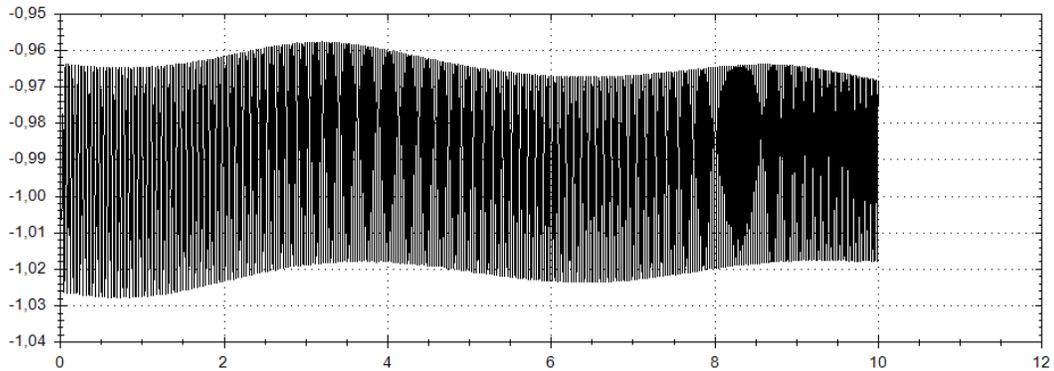

Fig.1. Results of numerical calculations of coordinate $x(f)$ by Eqns. (9) for Jupiter. We have chosen for our calculations here parameter as follows: $h = 4 \times 10^{-16}$.

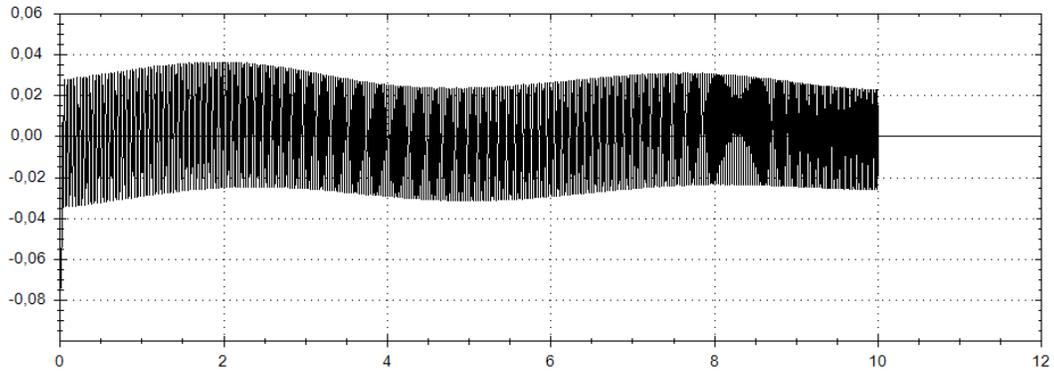

Fig.2. Results of numerical calculations of coordinate *y*(*f*) by Eqns. (9) for Jupiter.

We have chosen for our calculations here parameter as follows: $h = 4 \times 10^{-16}$.

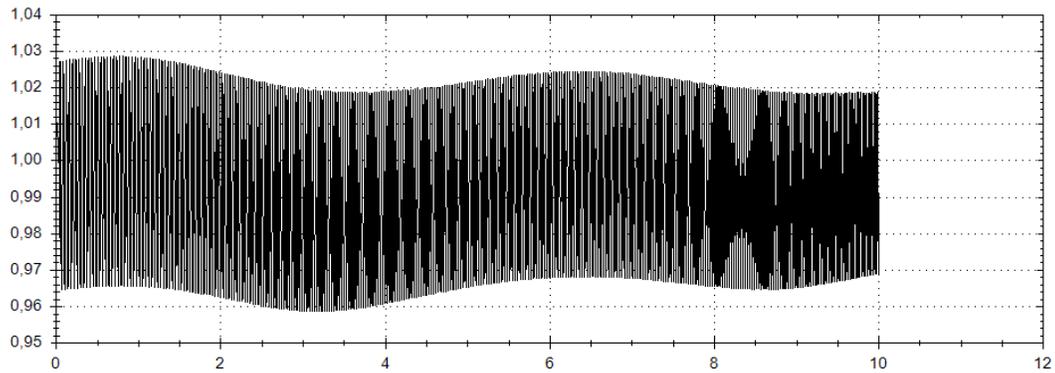

Fig.3. Results of numerical calculations of the distance $r_1$(*f*) by Eqns. (9) for Jupiter.

We have chosen for our calculations here parameter as follows: $h = 4 \times 10^{-16}$.

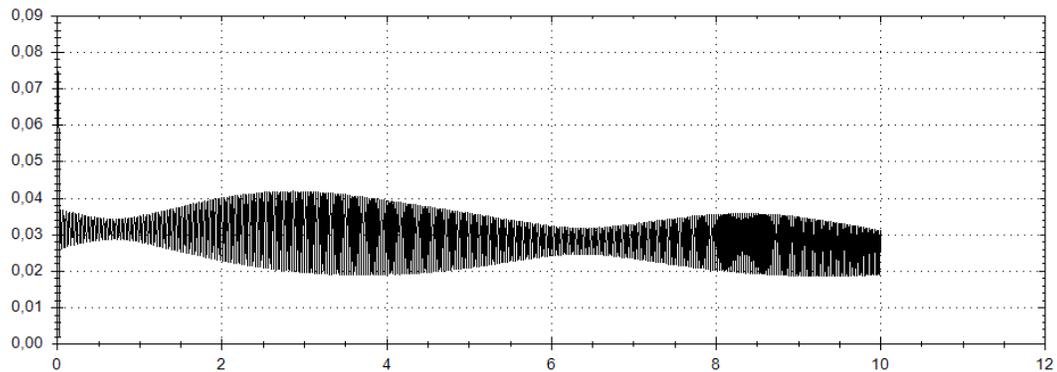

Fig.4. Results of numerical calculations of distance $r_2$(*f*) by Eqns. (9) for Jupiter.

We have chosen for our calculations here parameter as follows: $h = 4 \times 10^{-16}$.

As we can see from Figs.3-4, satellite experiences quasi-oscillations during its trapped motion around the planet revealing the obvious phenomenon of being captured into resonance with planet in this motion in ER3PB (case of two primaries for system "Jupiter-Sun").

Also, let us consider the additional case of two primaries for system "Earth-Sun" where eccentricity is chosen to be $e = 0.0167$, $\mu \cong 3.040 \times 10^{-6}$ in scheme of our numercial experiments presented on Figs. 5-8 (where $h$ is of same order as in case above). With help of Runge–Kutta fourth-order method, we have with step 0.001 starting from initial values: $x_0 = -1.021$, $(\dot{x})_0 = -0.35$, $y_0 = -0.082$, $(\dot{y})_0 = -0.33$ (Figs. 5-8) for the case of Earth.

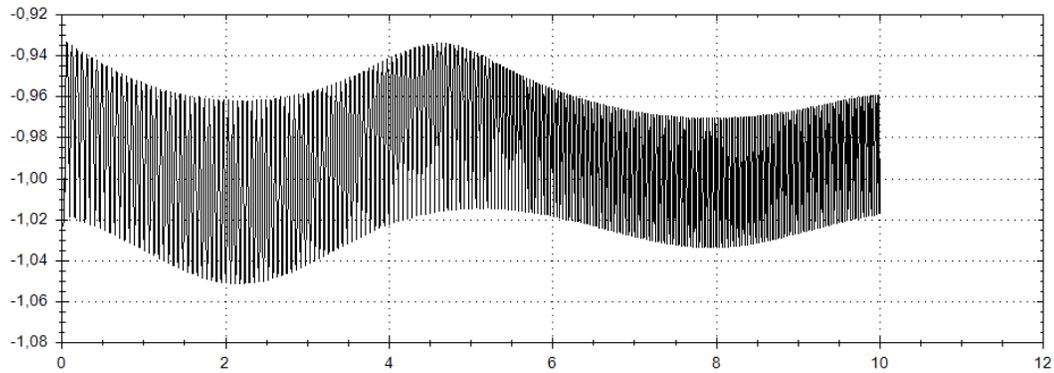

Fig.5. Results of numerical calculations of coordinate $x(f)$ by Eqns. (9) for Earth. We have chosen for our calculations here parameter as follows: $h = 4 \times 10^{-16}$.

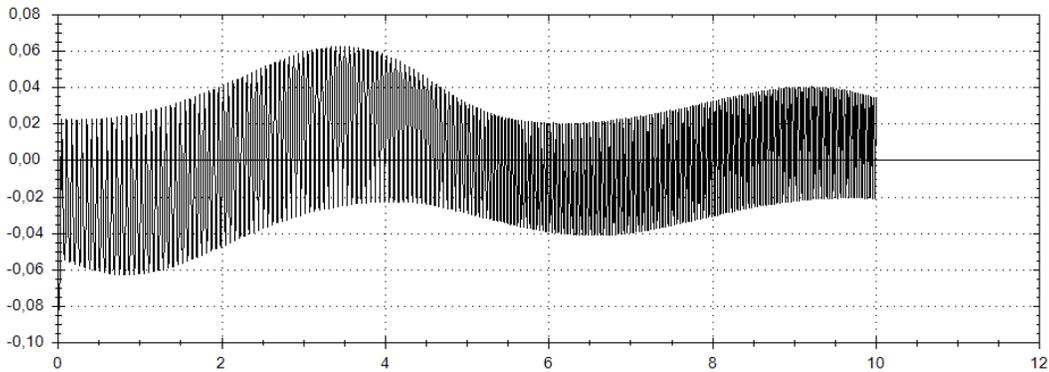

Fig.6. Results of numerical calculations of coordinate $y(f)$ by Eqns. (9) for Earth. We have chosen for our calculations here parameter as follows: $h = 4 \times 10^{-16}$.

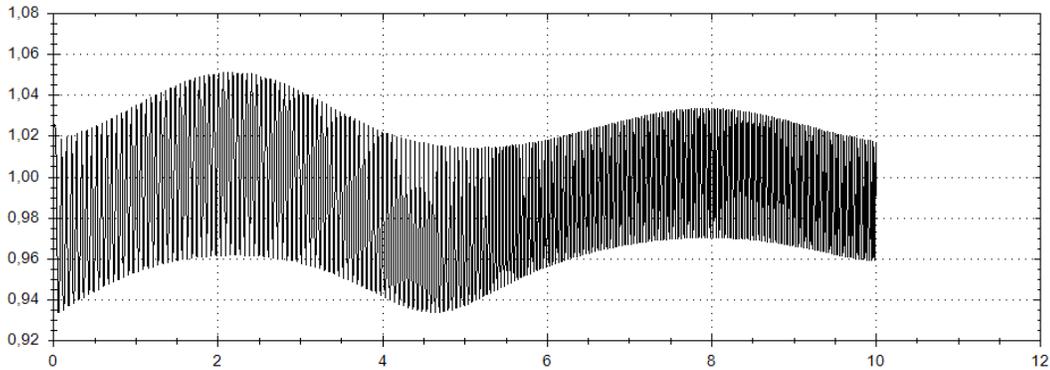

Fig.7. Results of numerical calculations of the distance $r_1(f)$ by Eqns. (9) for Earth. We have chosen for our calculations here parameter as follows: $h = 4 \times 10^{-16}$.

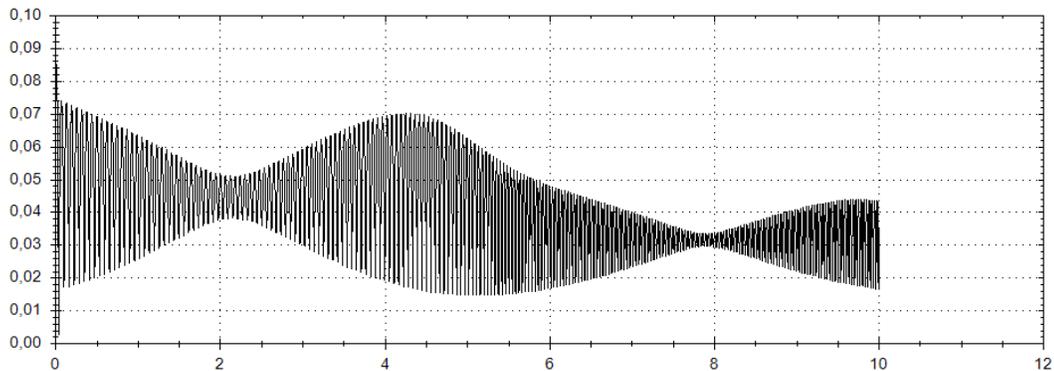

Fig.8. Results of numerical calculations of the distance $r_2(f)$ by Eqns. (9) for Earth. We have chosen for our calculations here parameter as follows: $h = 4 \times 10^{-16}$.

As we can obviously see from Figs.7-8, satellite also experiences quasi-oscillations during its trapped motion around the planet (here, Earth) revealing phenomenon of being captured into resonance with planet during such the type of motion in ER3PB (here, case of two primaries for system "Earth-Sun").

5. **Discussion & Conclusion.**

A lot of meaningful scientific contributions have been made in the field of Celestial Mechanics when discussing the obvious interconnection between dynamics of *finite-sized* satellite and its dependence on the gravitational influence from superposition of fields of both the primaries in ER3BP, elliptic restricted problem of three bodies.

A novel approach for solving equations of the dynamics of *finite-sized* satellite in case of the elliptic restricted three-body problem, ER3BP is presented in this work. We consider two primaries, $M_{Sun}$ and $m_{planet}$ (the last is secondary in that binary system), both are orbiting around their barycenter on *elliptic* orbits.

Our aim was to revisit previously presented in work [1] approach and to investigate the updated type of the dynamics of *finite-sized* satellite correlated implicitly to the motion of primaries (in the synodic co-rotating Cartesian coordinate system) in so way that satellite will always to be located near the secondary planet, $m_{planet}$, moving in this *trapped* motion on quasi-stable elliptic orbit.

We should note that the same results as presented here by solving Eqns. (7)-(9) (on Figs.1-8) could be achieved for the more general case of the satellite form with $k \neq 2h$ in (2). This will (possibly) bring the crucial changing in the character of dynamics of satellite with respect to the previous case $k = 2h$. We have presented here the most simplified (partial) case or sub-family of a wide class of solutions, whereas other cases could be investigated in the next researches (if required). Let us especially note that the distance of small mass $m$ (satellite) from second primary should exceed the level of minimal distances *out of double Roche-limit* for this chosen primary [5].

It is worth to note that we have succeded in numerical solving of system (7) (bearing (10) in mind) for system "Jupiter-Sun" where eccentricity is chosen to be $e = 0.048775$, $\mu \cong 954.509 \times 10^{-6}$ in the scheme of our numercial experiments presented on Figs. 9-12 (with estimation of $h$ of the same order as in all previous cases, see **Appendix, A1**). We should note that we have used for calculating the data the Runge–Kutta fourth-order method with step 0.001 starting from initial values for the case of Jupiter:

$$x_0 = -0.98, \ (\dot{x})_0 = -0.1, \ y_0 = -0.03, \ (\dot{y})_0 = 0.2.$$

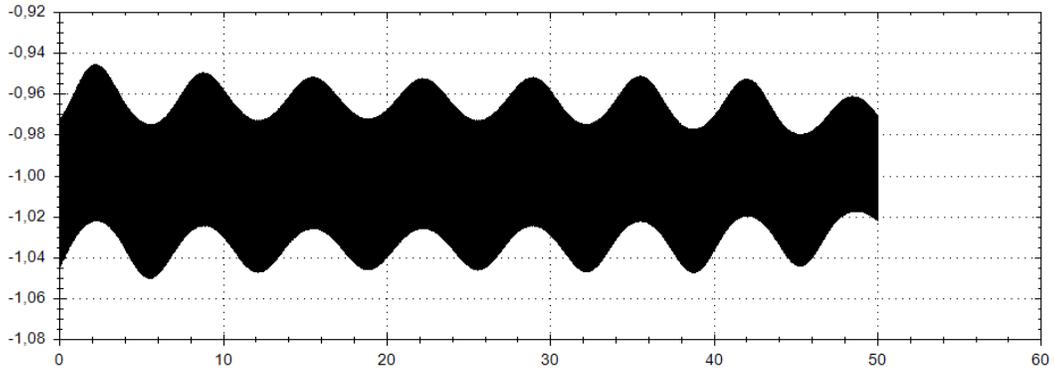

Fig.9. Results of numerical calculations of coordinate $x(f)$ by Eqns. (7) for Jupiter.

We have chosen for our calculations here parameter as follows: $h = 4 \times 10^{-16}$.

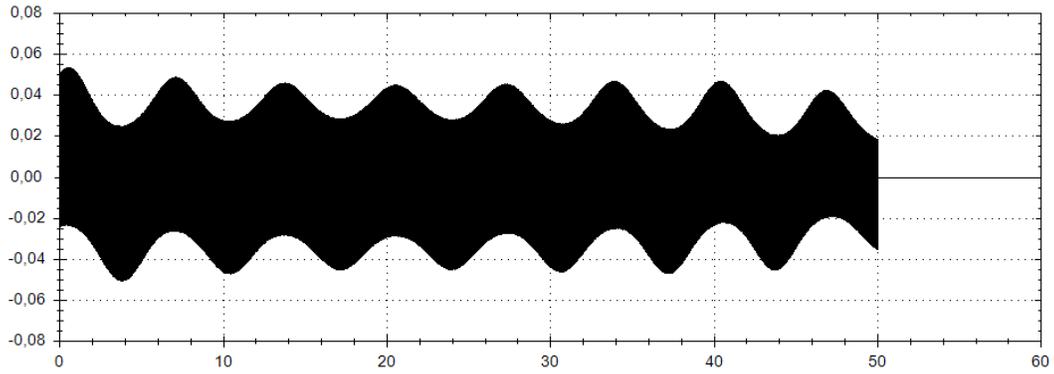

Fig.10. Results of numerical calculations of coordinate $y(f)$ by Eqns. (7) for Jupiter.

We have chosen for our calculations here parameter as follows: $h = 4 \times 10^{-16}$.

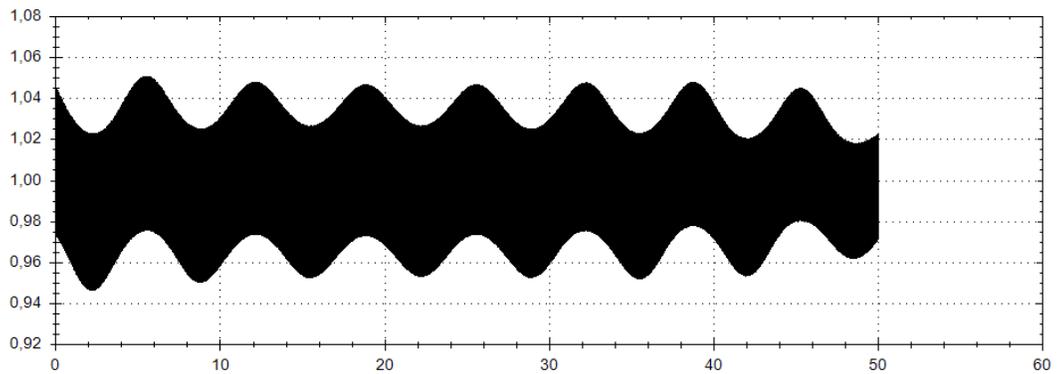

Fig.11. Results of numerical calculations of the distance $r_1(f)$ by Eqns. (7) for Jupiter.

We have chosen for our calculations here parameter as follows: $h = 4 \times 10^{-16}$.

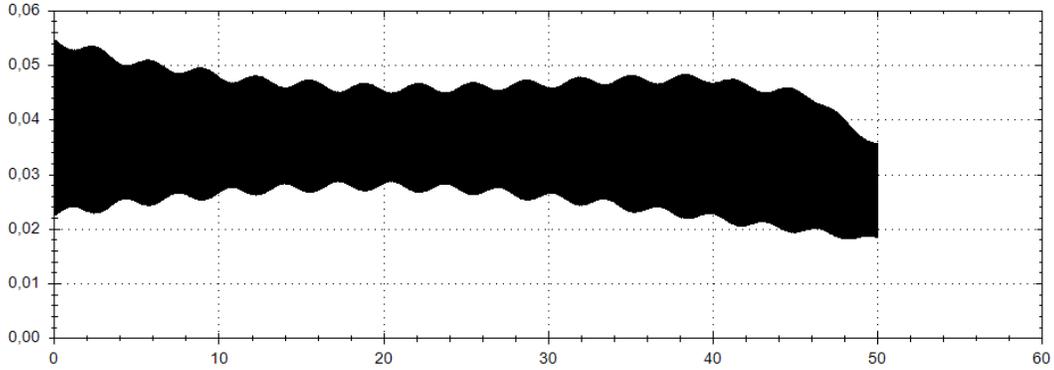

Fig.12. Results of numerical calculations of distance $r_2(f)$ by Eqns. (7) for Jupiter. We have chosen for our calculations here parameter as follows: $h = 4 \times 10^{-16}$.

As we can see from Figs.11-12, satellite experiences the quite often quasi-oscillating during its trapped motion around the planet revealing the obvious phenomenon of being captured into strong resonance with planet in this motion in ER3PB (case of two primaries for system "Jupiter-Sun").

Also, the remarkable articles should be cited, which concern the problem under consideration, [17-35]. As for geometrical presenting the coordinate system, used in the current research, it is worth to note that in [36] it was discussed that barycenter of Solar system is moving around center of Sun in its relative motion on quasi-periodic trajectory (less than $2.19\,R_{Sun}$ from center of Sun [36], where $R_{Sun}$ is the radius of Sun). An interesting fact is that the expression for potential at external point $\vec{r} = \{x, y, z\}$ for solid uniform ellipsoid of mass $m$ (which is supposed to be nearly spherical) in [37, p.300, see Eqn. (1.5) there], differs from that one which was the origin source [cited above] for expression (2) in the current work (pointed out in [3], see p. 111). This fact could be explained by various definitions in the used terms during the derivations of such the expressions in the aforementioned works (e.g., these expressions for potentials of ellipsoid have been mentioned with different signs in text of aforementioned researches), whereas coefficients in [3] were pointed out with specific details (in [37] them were pointed out in a general form).

**Appendix, A1 (estimation of absolute magnitudes for parameter *h* on Figs.1-12).**

Let us estimate the absolute magnitudes of parameter *h* which have been presented in calculations with their numerical solutions presented on Figs.1-12:

$$\frac{3h\,x^2}{10(x^2+y^2)} \ll \frac{|\vec{r}_2|}{|\vec{r}_1|} = \frac{((x-\mu+1)^2+y^2)^{\frac{1}{2}}}{((x-\mu)^2+y^2)^{\frac{1}{2}}} \ll 1$$

where we have chosen (e.g., on Figs.1-4) parameters as follows $e = 0.048775$, $\mu \cong 954.509 \times 10^{-6}$ for two primaries in system "Jupiter-Sun".

For the case of satellite Io of Jupiter (if we consider semimajor axis of Jupiter equals to 1), radius of Io equals to circa (1,821 km/(778,547,200 km)) = $2.339 \times 10^{-6}$, whereas order of *h* (proportional to the square of diameter) equals to circa $4 \times 10^{-16}$. We should note that the same estimation is valid for the case of Earth on Figs. 5-8 and of Jupiter on Figs. 9-12 (where *h* is of the same order as in case above).

Also, the same algorithm of scaled calculations for artificial satellite (which has size of order of circa 100 m) gives estimation circa $1.2 \times 10^{-24}$.

**Conflict of interest**

On behalf of all authors, the corresponding author states that there is no conflict of interest.

Remark regarding contributions of authors as below:

In this research, Dr. Sergey Ershkov is responsible for the general ansatz and the solving procedure, simple algebra manipulations, calculations, results of the article and also is responsible for the search of analytical and semi-analytical solutions.

Prof. Dmytro Leshchenko is responsible for theoretical investigations as well as for the deep survey in literature on the problem under consideration.

Dr. Alla Rachinskaya is responsible for obtaining numerical solutions related to approximated ones (including their graphical plots).

All authors agreed with results and conclusions of each other in Sections 1-4.